# Direct Observation of Chirality-Dependent Berry Curvature Dipole via Cross-Axial Nonlinear Hall Effect in Trigonal Te

*Jaehyun Lee*[1]‡, *Junhee Shin*[1]‡, *Daryll J. C. Dalayoan*[1], *Miju Park*[1], *Changhee Sohn*[1], *Gang Qiu*[2], *Noejung Park*[1], *Binghai Yan*[3], *Tony Low*[2], *Hosub Jin*[1]*, *Seon Namgung*[1]*

[1] Department of Physics, Ulsan National Institute of Science and Technology (UNIST), Ulsan, Republic of Korea

[2]Department of Electrical and Computer Engineering, University of Minnesota, Minneapolis, MN 55455, USA

[3]Department of Physics, Pennsylvania State University, University Park, PA 16802, USA

*Correspondence to: seon@unist.ac.kr, hsjin@unist.ac.kr

KEYWORDS: Nonlinear Hall Effect, Berry Curvature Dipole, Tellurium, Chirality.

Quantum geometric quantities, such as Berry curvature (BC) and quantum metric, are known to govern the physical and topological properties of materials, making access to these quantities critically important for the comprehensive understanding of the materials. Chiral crystals offer a natural platform in which structural handedness can control quantum geometry and nonlinear transport, but direct experimental access to the quantum geometric quantities has remained elusive. Here, we show that trigonal tellurium (Te) hosts an enantiomer-dependent Berry curvature dipole (BCD) that can be accessed through the measurement of cross-axial nonlinear Hall effect (NHE), unlike previously studied in-plane configurations where the BCD contribution vanishes. Angle-resolved measurements on left- and right-handed Te devices reveal a second-order transverse response whose sign reverses with crystal chirality. The extracted cross-axial nonlinear susceptibility, $\chi_{yxz}$, exceeds the in-plane susceptibility, $\chi_{xxx}$, by more than two-folds, indicating a strong BCD-driven contribution. First-principles calculations attribute this behavior to a topologically nontrivial hedgehog-like BC texture near the H points and predict a BCD that reverses sign between opposite enantiomers, consistent with experiment. These results establish cross-axial nonlinear Hall transport as a symmetry-selective probe of chirality-controlled quantum geometry in tellurium and related chiral materials.

## INTRODUCTION

Quantum geometric quantities in solids, such as quantum metric and Berry curvature (BC), provide a powerful framework for linking measurable transport and optical responses to the underlying electronic structure. For example, the Hall conductivity in the linear response regime is determined by the integral of Berry curvature over the Brillouin zone, reflecting its topological origin. Likewise, the shift current is governed by the shift vector, which originates from the Berry connection closely related to quantum geometry[1-9]. A comprehensive understanding of a material therefore requires identifying the relevant quantum geometric quantities that govern its physical response. This intimate connection should allow us to devise suitable experimental approaches for accessing quantum geometric quantities. For instance, the quantum metric tensor, which characterizes the distance between quantum states, has recently been measured using polarization-dependent angle-resolved photoemission spectroscopy (ARPES)[10]. Similarly, the shift current has been directly observed through terahertz electromagnetic spectroscopy in ferroelectric semiconductors[6]. Once established, these quantum geometric quantities can be further used to infer broader physical and topological properties[11-13] and concomitantly to guide the design of materials with tailored functionalities[14-16].

Along these lines, the nonlinear Hall effect (NHE) has emerged as a powerful probe of Berry curvature dipole (BCD)[17-21], which is one of the central quantum geometric quantities governing second-order transport. The BCD has been identified as a key quantum geometric quantity governing diverse quantum phenomena such as circular photogalvanic effect[22] and valley orbital magnetization[23]. Because the NHE can arise in inversion-symmetry-broken systems without requiring an external magnetic field or magnetic ordering, it provides a particularly powerful means to access the BCD. Owing to its strong sensitivity to inversion symmetry breaking, the NHE has also emerged as a promising probe of phase transitions in materials such as ferroelectric[24-26] and anti-ferromagnetic states[27]. Moreover, this method is especially valuable under extreme conditions, such as high magnetic fields or high pressure, where conventional optical probes including second-harmonic generation (SHG) are not readily applicable[28].

Trigonal Te, a nonmagnetic chiral semiconductor, has been extensively studied for its rich chirality-related phenomena, including nonreciprocal transport[29-31], negative magneto-resistance[32,33], Weyl-related electronic properties[34-36], and chirality induced spin selectivity (CISS) [37]. Due to the broken inversion symmetry in the mirror-symmetry broken system, Te is also expected to exhibit a NHE driven by BCD[38,39]. More importantly, the existence of two opposite enantiomers provides a unique symmetry knob for tuning the Berry curvature, which changes sign with crystal handedness. Te therefore offers an ideal setting in which to investigate the connection between quantum geometric quantities and measurable transport in a chiral solid (Figure 1). Beyond the fundamental interest, establishing this chirality-dependent NHE may enable device concepts such as opposite-direction rectifiers and provide a route to studying topological transitions at interfaces between opposite enantiomorphs[40].

Previous studies reported the NHE measurements in chiral Te[30,41-43]. However, those measurements did not directly access the BCD, because they were performed in in-plane configurations where the intrinsic BCD contribution vanishes by symmetry. In trigonal Te, the BCD tensor is diagonal, $D_{xx} = D_{yy} = -D_{zz}/2$, and the intrinsic nonlinear Hall response satisfies $\chi_{abc} \propto -\varepsilon_{adc} D_{bd}$. Since the Levi-Civita $\varepsilon_{adc} = 0$ if any two indices are the same, the BCD only contributes to the cross-axial susceptibility $\chi_{yxz}$, while the purely in-plane component $\chi_{xxx}$ is not generated by the intrinsic BCD. This makes cross-axial NHE measurements essential for directly probing the BCD in Te. In this work, we demonstrate an enantiomer-dependent BCD in trigonal Te through cross-axial NHE measurements. Our experimentally measured $\chi_{yxz}$ for the two enantiomers is more than two-folds larger than $\chi_{xxx}$, evidencing the dominant role of quantum geometry in the nonlinear transverse response. First-principles density functional theory (DFT) calculation clearly identifies a hedgehog-like Berry-curvature texture as the microscopic origin of the BCD and show that it reverses with crystal handedness, while resistivity scaling supports the dominance of the intrinsic BCD over extrinsic scattering mechanisms.

## RESULTS

**Berry curvature dipole and nonlinear Hall effect in chiral Te.** To access the BCD in trigonal Te, we focus on the intrinsic nonlinear Hall response that is symmetry-selectively coupled to the BCD. In the semiclassical description, the intrinsic BCD contribution to the second-order conductivity is given by[17],

$$\chi_{abc} = -\varepsilon_{adc} \frac{e^3\tau}{2(1+i\omega\tau)} \int_k f_0(\partial_b \Omega_d) \equiv -\varepsilon_{adc} \frac{e^3\tau}{2(1+i\omega\tau)} D_{bd} \quad (1)$$

where $\Omega_d$ is BC and $D_{bd}$ is the BCD tensor. As schematically illustrated in Figure 1, the Berry curvature of Te is odd under inversion of crystal chirality, so the corresponding BCD and nonlinear Hall response are likewise expected to reverse sign between opposite enantiomers. Chirality therefore provides a useful symmetry tuning parameter for identifying the connection between the BCD and the nonlinear Hall effect (NHE) more clearly. For trigonal Te, symmetry constrains the BCD tensor to a diagonal form, $D_{xx} = D_{yy} = -D_{zz}/2$, with all off-diagonal components vanishing (Supplementary Note 1). Consequently, the intrinsic BCD contributes to the cross-axial susceptibility $\chi_{yxz}$, whereas the purely in-plane component $\chi_{xxx}$ is not produced by the intrinsic BCD. This symmetry relation makes a cross-axial measurement essential for directly probing the BCD in Te.

Te is a nonmagnetic elemental semiconductor composed of helical atomic chains stacked along the threefold screw axis (*c*-axis), and it exists in two opposite enantiomeric forms, right-handed (RH) and left-handed (LH), as schematically shown in Figure 1. Single-crystalline Te flakes with thicknesses in several tens of nanometers and lateral sizes in tens of micrometers were synthesized using a hydrothermal process[44]. Te with two irreducible images is classified as enantiomorphic chiral space groups $P3_121$ or $P3_221$, corresponding to RH and LH Te respectively, confirmed by scanning transmission electron microscopy (STEM) [45] (Figure 2a). This enantiomer-resolved identification is essential for establishing the sign reversal of the BCD and the nonlinear Hall response with crystal chirality.

To probe the BCD of chiral Te by measuring the out-of-plane component of NHE, a 2D Te flake was transferred onto bottom electrodes with angular patterns, and then a top metal contact was made (Figure 1d and Figure 2b). The second-order susceptibility of NHE and the corresponding BCD can be determined by the angle-dependent NHE, because it is a function of the angle $\theta$ between the c-axis (parallel to $\hat{z}$) and the direction of current injected in the $\hat{x}-\hat{z}$ plane (Supplementary Note 2). Hexagonal boron nitride (h-BN) layers were inserted to electrically insulate most of the flake except for the exposed metal contact region with an area of 1 μm × 1 μm, allowing for the measurement of the second-order transverse voltage detected along the out-of-plane $\hat{y}$-direction.

Whereas the linear Hall response requires time-reversal symmetry breaking, the second-order nonlinear Hall response is allowed under time-reversal symmetry and instead relies on inversion symmetry breaking[17,28]. Since elemental trigonal tellurium lacks inversion symmetry, the nonlinear second-order transverse charge transport occurs even in the absence of magnetic field. In the weak field limit the current density is expressed as $j_i = \sigma_{ij}E_j + \chi_{ijk}E_jE_k$ with the linear conductivity $\sigma_{ij}$ and the second-order non-linear susceptibility $\chi_{ijk}$. The symmetry-allowed nonlinear susceptibility tensor for the space group of Te is expressed as follows[41].

$$\chi = \begin{pmatrix} \chi_{xxx} & \chi_{xyy} & 0 & \chi_{xyz} & 0 & 0 \\ 0 & 0 & 0 & 0 & \chi_{yxz} & \chi_{yxy} \\ 0 & 0 & 0 & 0 & 0 & 0 \end{pmatrix} \tag{2}$$

Here, the symmetry of Te enforces additional constraints of $\chi_{xxx}$ = -$\chi_{xyy}$ = -$\chi_{yxy}$ and $\chi_{yxz}$ = -$\chi_{xyz}$, and therefore only the two components of $\chi_{xxx}$ and $\chi_{yxz}$ are required to fully describe the second order response in chiral Te. Notably, eqation.1 indicates that the cross-axial component of $\chi_{yxz}$ is governed by the BCD, whereas the in-plane component of $\chi_{xxx}$ is not. The cross-axial component $\chi_{yxz}$ can be obtained by measuring the second-order transverse voltage in response to the in-plane electric fields of $\mathbf{E} = E^{\omega}(\hat{\boldsymbol{x}} \sin\theta + \hat{\boldsymbol{z}} \cos\theta)$, considering the equation of $j_y^{2\omega} = \chi_{yxz}(2|E^{\omega}|^2 \sin\theta\cos\theta)$. Experimentally, the value of $\chi_{yxz}$ can be obtained more accurately by fitting to multiple sets of $E_x^{\omega}$ and $E_z^{\omega}$, which are readily obtained by varying the angle between a straight electrode pair and the chiral axis (z-axis). Accordingly, the equation can be recast to show the relation between the second-order Hall voltage $V_{\perp}^{2\omega}$, the nonlinear susceptibility for a given a.c. current density $j^{\omega}$, and an angle $\theta$ between the current density and the z-axis in the plane (equation 3). The derivation of the equations including that of the longitudinal one ($\chi_{xxx}$) is provided in Supplementary Note 2.

$$\frac{V_{\perp,out}^{2\omega}}{(I^{\omega})^2} = 2\frac{R_{xx}R_{yy}R_{zz}}{{L_x}^2} A_y \chi_{yxz} \sin\theta\cos\theta \qquad (3)$$

Here, $A_y$ is the area of the vertical electrode and $L_x$ (= $L_z$) is the distance between electrode contacts along x(z)-axis.

**The cross-axial Nonlinear Hall effect in Te.** To obtain the second-order susceptibilities, an a.c. current $I^{\omega}$ (77.7 Hz) was injected between an in-plane electrode pair aligned at an angle $\theta$ from the z-axis (c-axis), while the second order transverse voltage was measured either along the *out-of-plane* or *in-plane* direction (Figure 2b). The *out-of-plane* transverse voltage $V_{\perp,out}^{2\omega}$ (i.e. along y-axis) for different $\theta$ values in the LH and the RH Te device is shown in Figure 2c and 2d, respectively. In both devices, the NHE voltages scale linearly with $(I^{\omega})^2$ for all measured angles, confirming its second-order origin. Figure 2e summarizes the angular dependence of the ratio of the NHE voltages to the square of the input current (i.e. $V_{\perp,out}^{2\omega}/(I^{\omega})^2$) for out-of-plane configuration, respectively, and those are well fitted with the equation (3). Notably, the NHE response reverses sign between the two enantiomers, consistent with their opposite crystal handedness and indicating that the underlying mechanism is chirality dependent. By fitting the observed data to the equation (3), the corresponding non-linear susceptibilities are obtained for both Te devices with opposite chirality as shown in Table 1. The magnitude of the susceptibility is similar for Te devices with different chirality, but their signs are opposite. For comparison, the in-plane NHE was also measured as shown in Supplementary Figure 1, and the corresponding longitudinal susceptibility ($\chi_{xxx}$) governing *in-plane* NHE is determined (Supplementary Table 1). The ratio between the second-order susceptibility governing *out-of-plane* and *in-plane* NHE (i.e. $|\chi_{yxz}|/|\chi_{xxx}|$) is 2.28 and 2.81 for the LH-Te device and RH-Te device, respectively. In other words, the cross-axial susceptibility linked to the BCD is significantly larger than the in-plane one, highlighting the important role of quantum geometry in the nonlinear transverse response of trigonal Te. Given that Te already exhibits a giant in-plane nonlinear Hall response compared with many other materials[42], the even larger cross-axial susceptibility further underscores its promise for applications such as rectification devices.

**DFT calculation of Berry curvature of Te.** To further clarify the microscopic origin of the measured cross-axial nonlinear Hall effect, the BC and its dipole were calculated using DFT. In the electronic structure of chiral Te, the conduction band minimum and valence band maximum are located near the H-point, leading to the dominant BC texture in its vicinity (Figure 3a). Notably, due to the helical geometry of bulk Te, a topologically non-trivial hedgehog-type BC distribution appears. Along the H-K line, the in-plane components of the BC vanish under the three-fold screw rotation symmetry ($C_{3z}$). In contrast, under two-fold rotation symmetry ($C_{2x}$), the component parallel to the rotational axis remains along the A-H and H-L lines. In the left-handed Te, the BC is oriented toward the H point within the $k_x$–$k_y$ plane and is directed outward from the H point along the $k_z$ line (Figure 3b). In the 1st Brillouin zone of Te, there are two distinct H and $\mathrm{H}'$ points which are connected by the time-reversal symmetry. The topological nature of BC is preserved for the two time-reversal partners, resulting in identical hedgehog textures (*i.e.* $\vec{\Omega}(\vec{\mathrm{H}}+\vec{q}) = -\vec{\Omega}(-\vec{\mathrm{H}}-\vec{q}) = -\vec{\Omega}(\vec{\mathrm{H}'}-\vec{q})$ ). Furthermore, the BC distribution changes from hedgehog to anti-hedgehog in right-handed chiral Te. From the hedgehog BC distribution in Figure 3b, we plotted the 3-dimensional vectors $(\partial_x\Omega_x, \partial_y\Omega_y, \partial_z\Omega_z)$ which are responsible for the diagonal components of the BCD. $(\partial_x\Omega_x, \partial_y\Omega_y, \partial_z\Omega_z)$ near the H point are aligned along the $(-,-,+)$ direction in the left-handed Te, whereas the configuration is reversed to $(+,+,-)$ in the right-handed counterpart (Figure 3c). The BCD, given by $D_{ab} = \int_{\mathrm{BZ}} f_0\, \partial_a\, \Omega_b d^3k$, is shown in Figure 3d for both left- and right-handed Te. Based on the symmetry analysis, the diagonal components of the BCD are nonzero with $D_{xx} = D_{yy} = -D_{zz}/2$, and all off-diagonal components vanish (see Supplementary Note 1). The BCD also reverses its sign for opposite handedness, consistent with the experimentally observed chirality-dependent cross-axial NHE. These results demonstrate that the cross-axial NHE is directly linked to the BCD.

**Experimental estimation of BCD.** In addition to the theoretical calculation, we also estimated the BCD from the experimentally measured nonlinear Hall voltages for both devices with opposite chirality. The intrinsic BCD ($D_{xx}$) contribution to the non-linear susceptibility tensor is expressed as $\chi_{yxz} = -\varepsilon_{yxz}\frac{e^3\tau}{2\hbar^2(1+i\omega\tau)}D_{xx}$ with $j_y^{2\omega} = \sigma_{yy}E_y^{2\omega} = 2\chi_{yxz}E_x^{\omega}E_z^{\omega}$, which yields equation (4), in the DC limit $\omega\tau \ll 1$(see Supplementary notes 3).

$$\frac{E_y^{2\omega}}{E_x^{\omega} E_z^{\omega}} = \frac{1}{\sigma_{yy}} \frac{e^3 \tau}{\hbar^2} D_{xx} \tag{4}$$

Standard one-band model was used to fit the linear Hall to extract the mobility of the devices, which further provided the scattering time $\tau$ (Supplementary notes 3). We obtained the scattering time $\tau = \frac{\mu m^*}{e} =$ 294 fs and 308 fs for RH- and LH- Te, respectively. Here, we used the conductivity extracted from the linear Hall measurement (Supplementary Figure 2) and the effective mass of charge carriers $m^* = 0.150m_0$, extracted from the band structure (Figure 3a). As a result, we obtained $D_{xx} = -0.0676$ and $D_{xx} = 0.0616$ from the RH- and LH-Te, respectively, which are within the range expected from the theoretical calculation shown in Figure 3c.

## DISCUSSION

The NHE has been known to be governed by BCD along with extrinsic factors of disorder-based scattering processes, such as skew-scattering and side-jump[46,47]. Even though precise quantification of each factor's contribution to the NHE is challenging[46,47], they can still be reasonably estimated by scaling the cross-axial nonlinear transverse response in terms of linear longitudinal conductivity (or resistivity) as follows[46]. Since resistivity varies with temperature, we can investigate the resistivity-dependent nonlinear Hall response by measuring both resistivity and the nonlinear Hall response across a range of temperatures. The temperature-dependent longitudinal resistivity $\rho_l$ was measured for the LH-Te device (Figure 4a) and the RH-Te device (Figure 4c). Along the measured cross-axial NHE, it was obtained the relation between the ratio of the NHE voltages to the square of the input current ($V_{\perp,out}^{2\omega}/(I^{\omega})^2$) and the longitudinal resistivity (Figure 4b and 4d). The data were fitted using the scaling form

$$\frac{V_{\perp,out}^{2\omega}}{(I^{\omega})^2} = \beta + \gamma\rho_l + \delta{\rho_l}^2 \tag{5}$$

where $\beta$, $\gamma$, and $\delta$ are scaling coefficients[41]. Within the scaling framework, the fitting parameters can be written as

$$\beta = \rho_{l0} + \left(C_{00}^{sk.1} - C_{01}^{sk.1} + C_{11}^{sk.1}\right){\rho_{l0}}^2,$$

$$\gamma = \left(C_0^{sj} - C_1^{sj} + C_{01}^{sk,1} - 2C_{11}^{sk,1}\right)\rho_{l0}, \tag{6}$$

$$\delta = C^{in} + C_1^{sj} + C_{11}^{sk,1},$$

where $\rho_{l0}$, $C^{in}$, $C^{sk}$, and $C^{sj}$ denotes the residual resistivity due to static impurities at zero temperature, the intrinsic BCD contribution, the skew-scattering contribution, and the side-jump contribution, respectively. The notations and the derivation of the equations are provided in Supplementary Notes 4, and the fitted coefficients are provided in Table 2. The sign-reversal of the full coefficient set upon switching crystal chirality is consistent with the result that the non-linear susceptibility tensor changes sign under inversion of the crystal chirality. The fitted coefficients provide important clues to determine which mechanisms dominantly contribute to the observed nonlinear transverse transport. Since $C^{in}$ is only incorporated in the coefficient δ, the contribution of BCD can be estimated by comparing the magnitude of $C^{in}$ with those of the other components of the coefficient δ. Here, we note the other components of $C_1^{sj}$ and $C_{11}^{sk,1}$ in the coefficient δ are also included in the coefficient γ and the signs of δ and γ are the same, as provided in Table 2. According to the equations (6), if the side-jump contribution $C_1^{sj}$ is assumed dominant, $\gamma$ and $\delta$ would be expected to have opposite signs because the sign of the $C_1^{sj}$ in the expression of $\gamma$ and $\delta$ is opposite. However, this is clearly inconsistent with the experimental results shown in Table 2 that $\gamma$ and $\delta$ have the same sign for each chirality. The same argument applies for the skew-scattering term of $C_{11}^{sk,1}$. Thus, it can be concluded that the intrinsic BCD contribution for the cross-axial nonlinear transverse transport of Te is dominant over the contribution from the side-jump or the skew-scattering mechanism. In contrast, however, $\gamma$ and $\delta$ have opposite sign for the *in-plane* configuration, where the NHE is entirely independent of the BCD and is rather governed by the side-jump mechanism[41] (Supplementary Table 2 and Supplementary Figure 3).

## CONCLUSION

In summary, we have shown that trigonal Te hosts an enantiomer-dependent BCD that can be accessed through cross-axial nonlinear Hall measurements. The opposite handedness of the crystal provides a natural symmetry knob, and the observed sign reversal of the cross-axial nonlinear Hall response between left- and right-handed Te directly reveals the chirality dependence of the BCD. The cross-axial second-order susceptibility is experimentally measured 2.28 and 2.81 times larger than the longitudinal one for the LH-Te device and the RH-Te device, respectively, highlighting the central role of Berry-curvature-induced quantum geometry in the nonlinear transverse response. The BCD of $D_{xx}$ from the hedgehog BC is experimentally determined from the cross-axial second-order susceptibility for both handedness and lies within a range comparable to theoretical calculation. The dominant contribution of BCD is also supported by the scaling behavior of $V_{\perp,out}^{2\omega}/(I^{\omega})^2$ depending on the resistivity. These results establish trigonal Te as a model platform for probing chirality-controlled quantum geometry and point to new opportunities for chirality-sensitive nonlinear devices, including rectification functionalities based on opposite enantiomers.

METHODS

**Chemical Synthesis of Te Flakes.** 105 mg of $Na_2TeO_3$ and 550 mg of poly(vinylpyrrolidone) (average molecular weight, 360,000) were dissolved in 30 mL of Milli-Q water. 3.65 mL of $NH_4OH$ solution (25% by weight in water) and 1.94 mL of hydrazine hydrate (80% w/w) were added while stirring. The solution was sealed in an autoclave and heated at 180 °C for 30 h. The resulting solution was precipitated by centrifugation (3500 rpm, 5 min) and washed three times with distilled water.

**Sample Fabrication.** The device was fabricated by the Dry transfer method. The h-BN and Te flake were sequentially transferred on top of each other to form the device structure. First, the bottom electrode was fabricated via e-beam lithography. An h-BN layer was transferred on to the bottom electrode with tip of the electrode exposed. The spider web configuration-electrodes were then fabricated on top of this structure by e-beam lithography. The Te flakes were then transferred onto this structure followed the top h-BN layer. Finally, the top electrode was then fabricated on top of the sensing region. 3 nm Ti / 30 nm Au were deposited using e-beam deposition for all electrodes.

**Transport Measurements.** The low-temperature transport measurement was performed by using the Cryogen-Free Measurement System (Cryogenic Ltd) at 10 K. The standard Hall effect was measured in van-der Pauw configuration at 10 K under an external out-of-plane magnetic field. For the second order NHE measurements, an a.c. current with frequency 77.7 Hz was applied to the electrodes of the desired angles of the device. Second order voltages $V_{\perp,in}^{2\omega}$ and $V_{\perp,out}^{2\omega}$ were measured at the second order frequency using an SR830 lock-in amplifier (Stanford).

**Structural Characterization.** Te flake lamellae were prepared using Ga+ ion beam milling using a Helios NanoLab 450HP FIB-SEM system (Thermo Fisher Scientific) with operating voltage 30 kV. The STEM images were acquired with a JEM-2100F microscope (JEOL Ltd.) in STEM mode. For the handedness determination, STEM images along the 010 zone axis were rotated 30° for mirrored images and compared with known chiral atomic models.

**Calculation on Band Structure and Berry Curvature Dipole of Chiral Te.** The electronic structure calculations were performed using projector-augmented wave (PAW) method and plane-wave basis set as implemented in VASP code package[48,49]. Exchange-correlation functional was treated with hybrid functional (HSE06)[50]. The potential was generated with the valence electron configurations: Te $5s^2 5p^4$. The energy convergence criterion for the self-consistent field approach was set to $10^{-6}$ eV. A plane-wave cutoff was 350 eV. 7×7×5 Monkhorst-Pack k-point mesh was chosen[51]. Spin-orbit coupling was included in the calculations. To calculate Berry curvature, a Wannier-interpolated tight-binding Hamiltonian including Te 5p-orbitals was constructed from the ab initio calculation, using the WANNIER90 code[52,53]. The TB Hamiltonian was symmetrized using the WannSymm code[54]. The Berry curvature and Berry curvature dipole was calculated in symmetrized TB model with a 800×800×800 k-mesh.

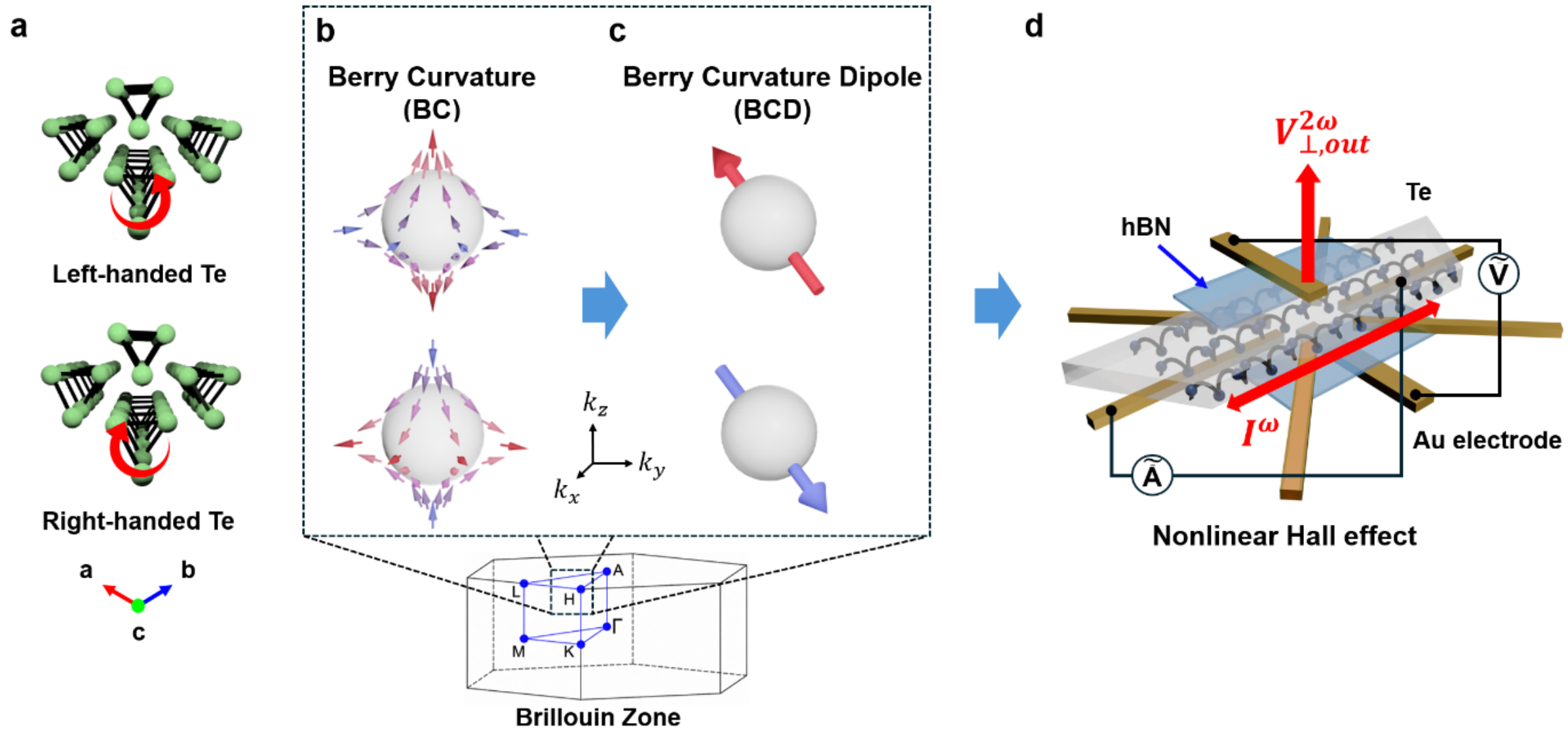

**Figure 1. Chirality-dependent Berry curvature dipole (BCD) of Te probed by cross-axial nonlinear Hall effect (NHE).** Derivatives of hedgehog Berry curvature (BC) dependent on the chirality are probed by cross-axial measurement of nonlinear Hall effect (NHE). The hedgehog BC and corresponding NHE are odd under inversion of crystal chirality of Te crystal. (a) Trigonal crystalline structure of Te with different chirality. (b, c) BC and BCD of Te. (d) Schematic device structure for measuring NHE of Te originated from BCD.

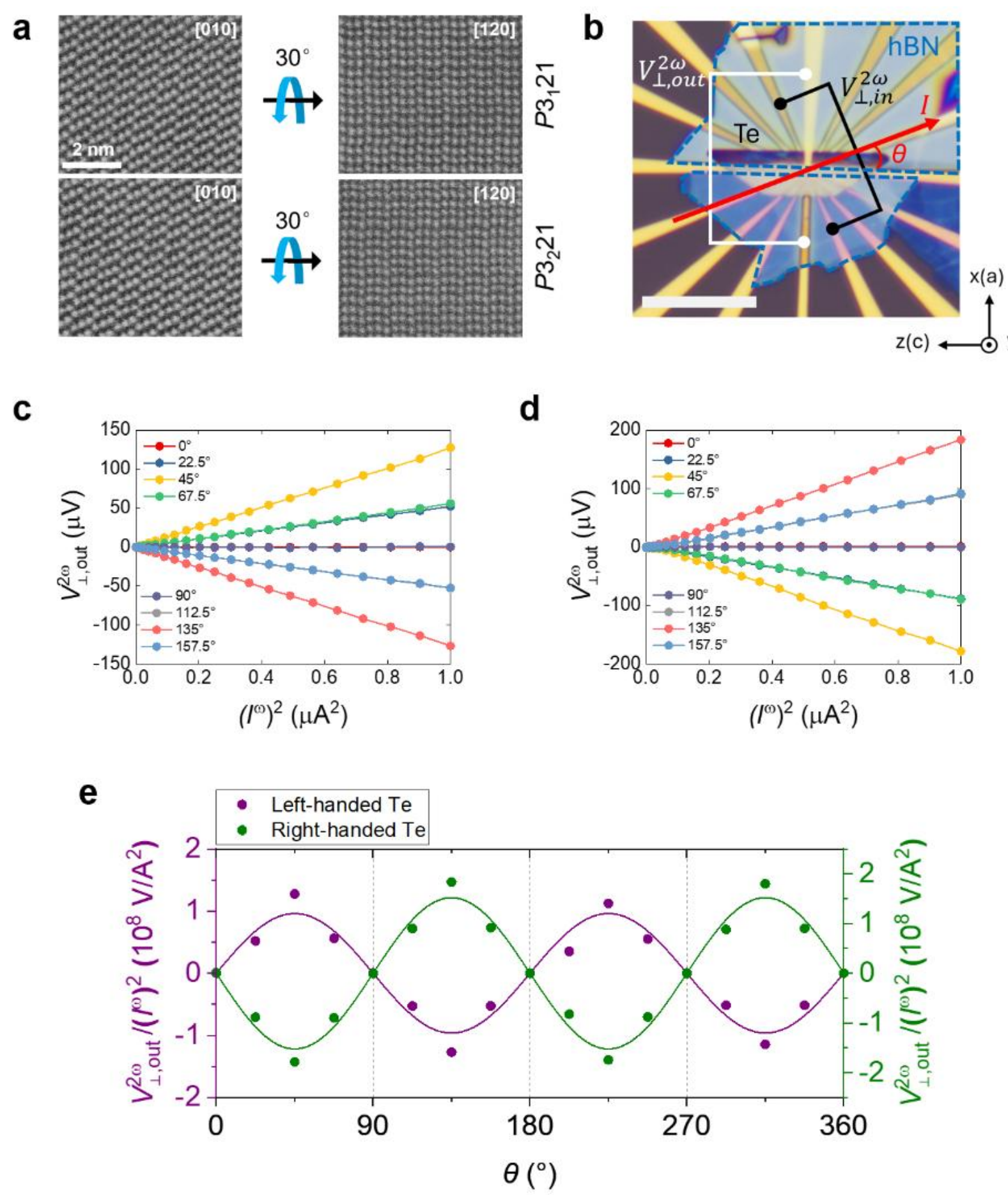

**Figure 2. Angle-dependent cross-axial nonlinear Hall effect in chiral Te devices.** (a) STEM images of two Te flakes with opposite chirality, showing 010 and 120 zone axis of the right-handed (top) and left-handed (bottom) Te. The handedness is confirmed by the rotation of 30°. (b) Optical image of a Te device. (c, d) *Out-of-plane* transverse (i.e. cross-axial) second order voltage $V^{2\omega}_{\perp,out}$ as a function of $(I^{\omega})^2$ at different angles of a left-handed and a right-handed Te device, respectively. (e) The ratio of the cross-axial NHE voltages to the square of the input current (i.e. $V^{2\omega}_{\perp,out}/(I^{\omega})^2$) as a function of $\theta$. The results of the left-handed (purple) and the right-handed (green) Te devices are shown in the plot together. The solid lines represent fits to the experimental data using the equation (3). The results reverse sign for opposite chirality.

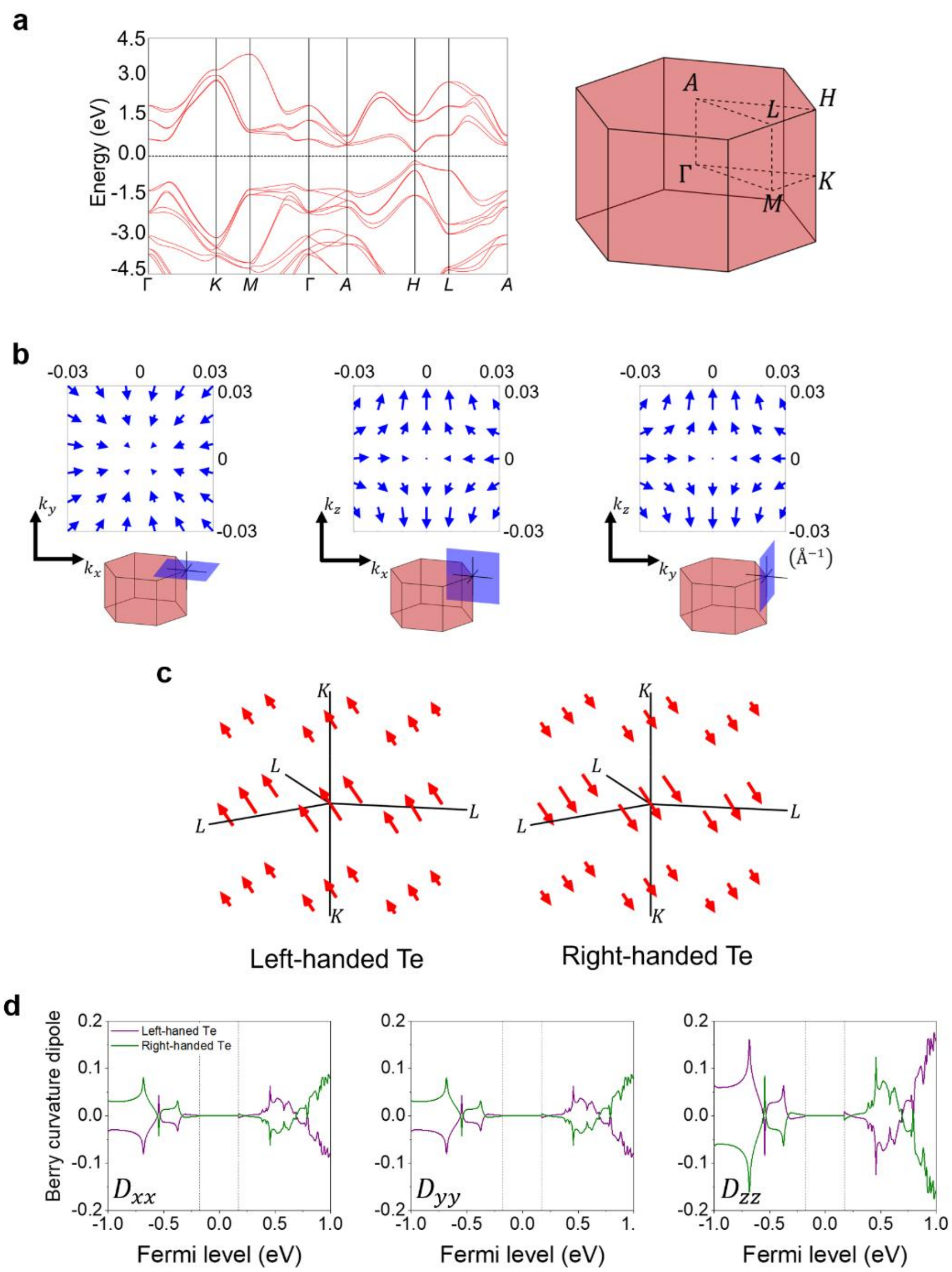

**Figure 3. Band structure, hedgehog Berry curvature texture, and Berry curvature dipole in trigonal Te.** (a) Band structure of the trigonal Te. (b) Berry curvature distribution of the left-handed chiral Te in the Brillouin zone around the H point in the $k_x$-$k_y$, $k_x$-$k_z$, and $k_y$-$k_z$ planes. (c) Derivatives of Berry curvature $(\partial_x\Omega_x, \partial_y\Omega_y, \partial_z\Omega_z)$ for left- and right-handed Te near the H point. (d) $D_{xx}$, $D_{yy}$, and $D_{zz}$ components of the Berry curvature dipole as a function of the Fermi level. Green and purple curves represent right- and left-handed Te, respectively.

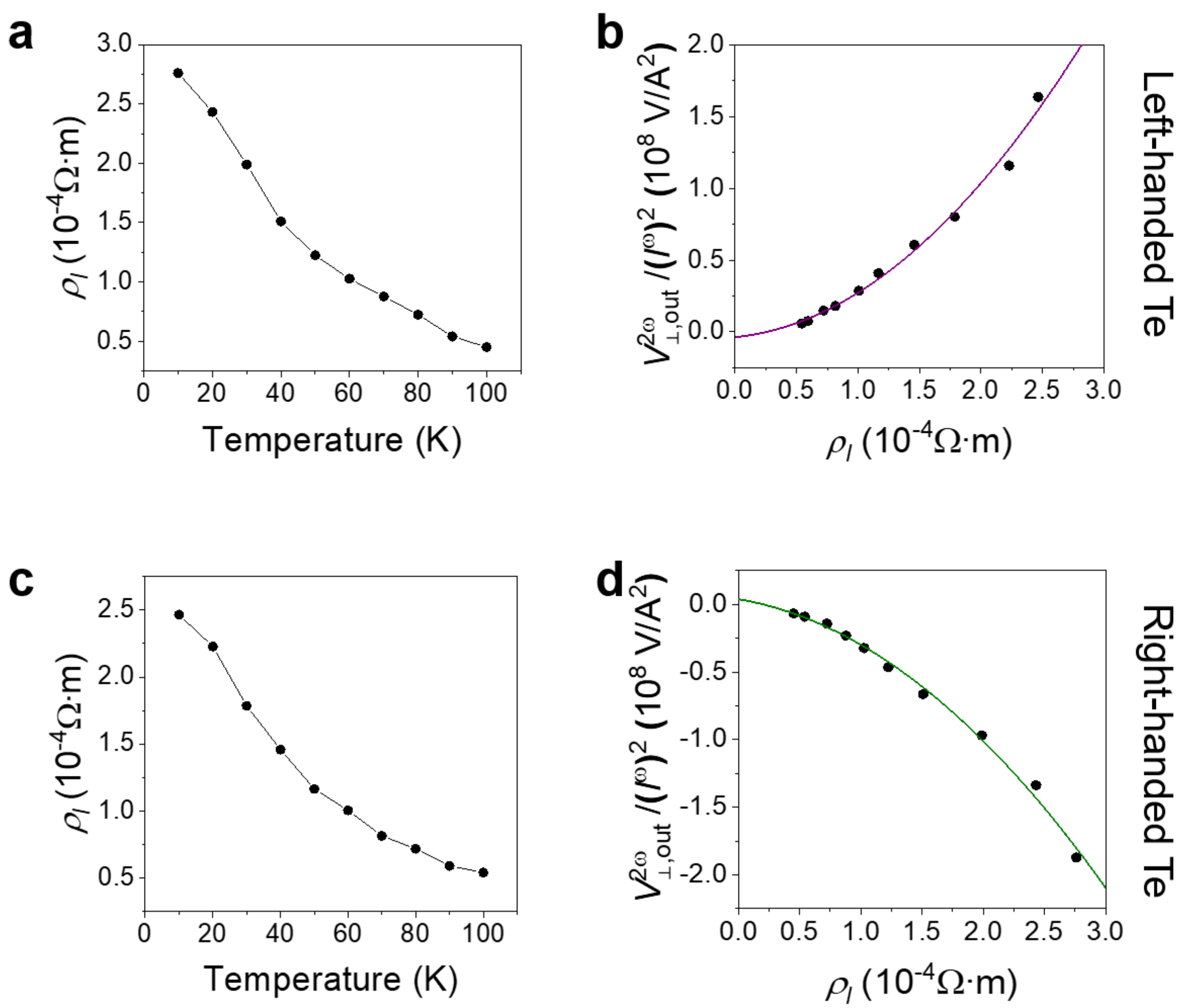

**Figure 4. Scaling of cross-axial nonlinear Hall effect as a function of resistivity.** (a, c) Temperature dependent longitudinal resistivity $\rho_l$ of a left- and right-handed Te device with the injected current $I^{\omega}$ at θ = 45˚, respectively. (b, d) The ratio of the cross-axial NHE voltages to the square of the input current (i.e. $V^{2\omega}_{\perp,out}/(I^{\omega})^2$) as a function of $\rho_l$ for the left- and right-handed Te device, respectively. The solid lines represent fits to the experimental data using the equation (5).

**Table 1.** The nonlinear susceptibility $\chi_{yxz}$ of left- and right-handed Te devices and corresponding BCD.

| | $\chi_{yxz}$ | $D_{xx}(=D_{yy})$ |
|---|---|---|
| Left-handed Te Device (Device 1) | $5.27\times10^{-3}/\Omega\cdot V$ | 0.0616 |
| Right-handed Te Device (Device 2) | $-6.26\times10^{-3}/\Omega\cdot V$ | -0.0676 |

**Table 2.** Fitted coefficients in equation (5) for scaling cross-axial nonlinear Hall effect with longitudinal resistivity for each chirality.

| | $\beta$ | $\gamma$ | $\delta$ |
|---|---|---|---|
| Out-of-plane, LH | -0.36 ($10^3$ V·A$^{-2}$) | 0.83 ($10^7$ m$^{-1}$·A$^{-1}$) | 2.27 ($10^{11}$ m$^{-2}$·V$^{-1}$) |
| Out-of-plane, RH | 0.38 ($10^3$ V·A$^{-2}$) | -1.49 ($10^7$ m$^{-1}$·A$^{-1}$) | -1.87 ($10^{11}$ m$^{-2}$·V$^{-1}$) |

ASSOCIATED CONTENT

**Supporting Information**. The following files are available free of charge.

Figure S1: Angle-dependent in-plane nonlinear Hall effect in Te; Figure S2: Linear Hall measurement of a Te device; Figure S3: Scaling of cross-axial nonlinear Hall effect as a function of resistivity; Table S1; The nonlinear susceptibility $\chi_{xxx}$ of left- and right-handed Te devices; Table S2; Fitted coefficients in equation (5) in the main text for scaling in-plane nonlinear Hall effect with longitudinal resistivity for each chirality; Note S1; Symmetry analysis of the BCD; Note S2: Non-linear Susceptibility $\chi_{yxz}$ and $\chi_{xxx}$; Note S3: Calculation of the Berry Curvature Dipole; Note S2: Scaling Law for the NHE of Te.

AUTHOR INFORMATION

**Corresponding Author**

*Seon Namgung, *Hosub Jin

**Author Contributions**

The manuscript was written through contributions of all authors. All authors have given approval to the final version of the manuscript. ‡These authors contributed equally.

ACKNOWLEDGMENT

This work was supported by the National Research Foundation of Korea (NRF) funded by the Korea government (MIST) under Grant No. RS-2024-00487988, RS-2022-NR069722, and RS-2023-NR076899. We thank UNIST Office of Research Facilities and Training (ResFacT) for the support of its facilities and equipment. We also thank Prof. Dai-Sik Kim (UNIST) for the support of equipment.

**Supplementary Information**

# Direct Observation of Chirality-Dependent Berry Curvature Dipole via Cross-Axial Nonlinear Hall Effect in Trigonal Te

Jaehyun Lee[1†], Junhee Shin[1†], Daryll J. C. Dalayoan[1], Miju Park[1], Changhee Sohn[1], Gang Qiu[2], Noejung Park[1], Binghai Yan[3], Tony Low[2], Hosub Jin[1*], Seon Namgung[1*]

[1]Department of Physics, Ulsan National Institute of Science and Technology (UNIST), Ulsan, 44919, Republic of Korea

[2]Department of Electrical and Computer Engineering, University of Minnesota, Minneapolis, 55455, USA

[3] Department of Physics, Pennsylvania State University, PA 16801, USA

*Correspondence to: seon@unist.ac.kr, hsjin@unist.ac.kr

[†]Contributed equally.

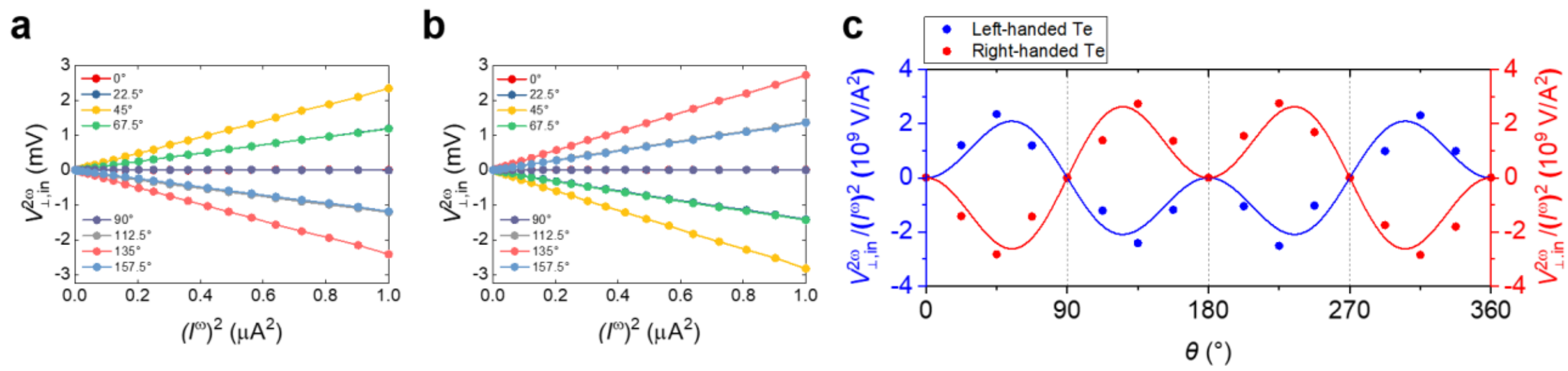

**Supplementary Figure 1. Angle-dependent in-plane nonlinear Hall effect in Te.** (a, b) *In-plane* transverse second order voltage $V_{\perp,in}^{2\omega}$ voltage as a function of $(I^{\omega})^2$ at different angles of a left-handed and a right-handed Te device, respectively. (c) The ratio of the NHE voltages to the square of the input current (i.e. $V_{\perp,in}^{2\omega}/(I^{\omega})^2$) as a function of $\theta$. The results of the left-handed (blue) and right-handed (red) Te devices are shown in the plot together. The solid lines represent fits to the experimental data using the equation shown in the supplementary note 2.

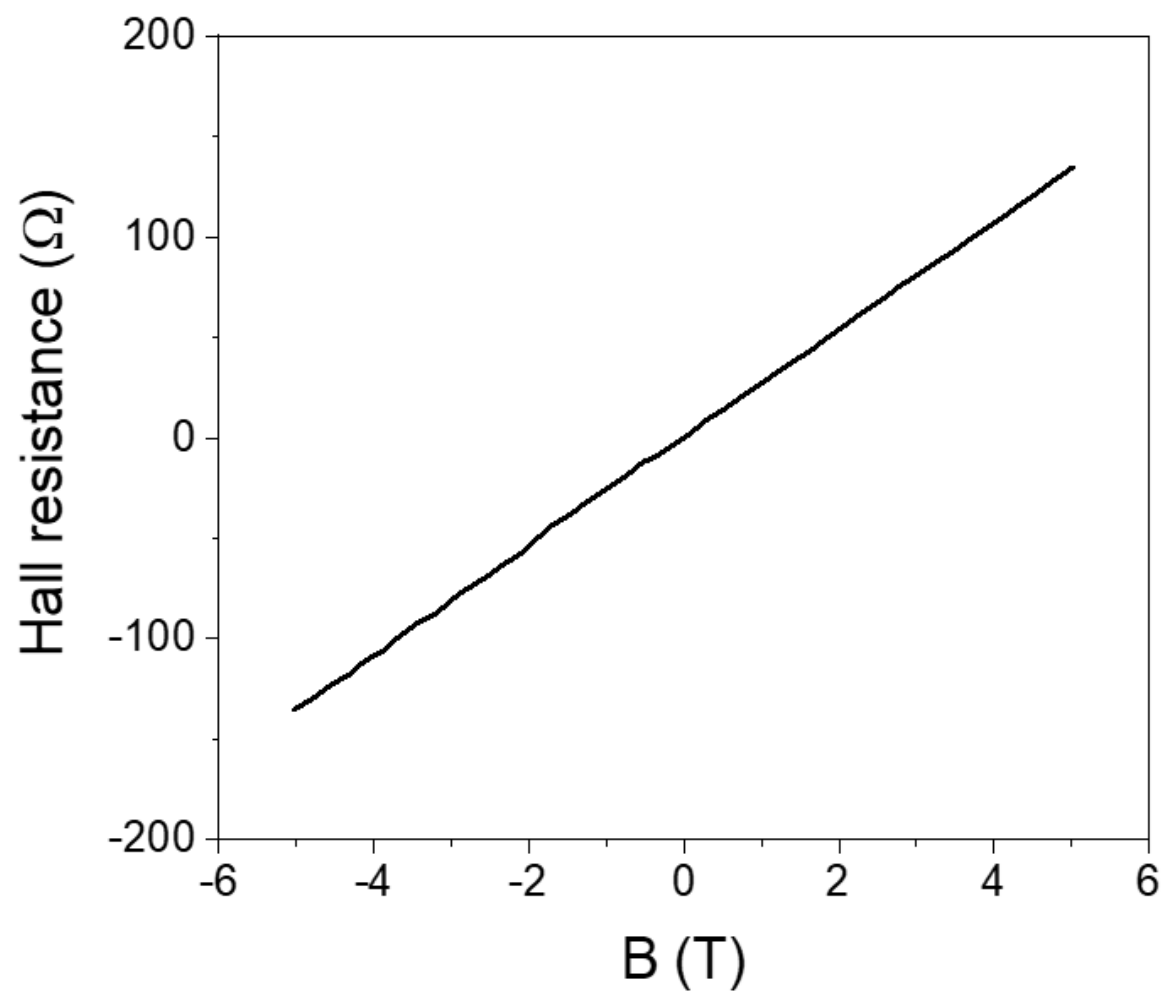

**Supplementary Figure 2. Linear Hall measurement of a Te device.** Hall resistance of a p-type Te measured as a function of magnetic field (B) -5 to 5 T at 10 K, showing a linear Hall response. The extracted carrier mobility is $\mu = 3444.73\ cm^2/V\cdot s$.

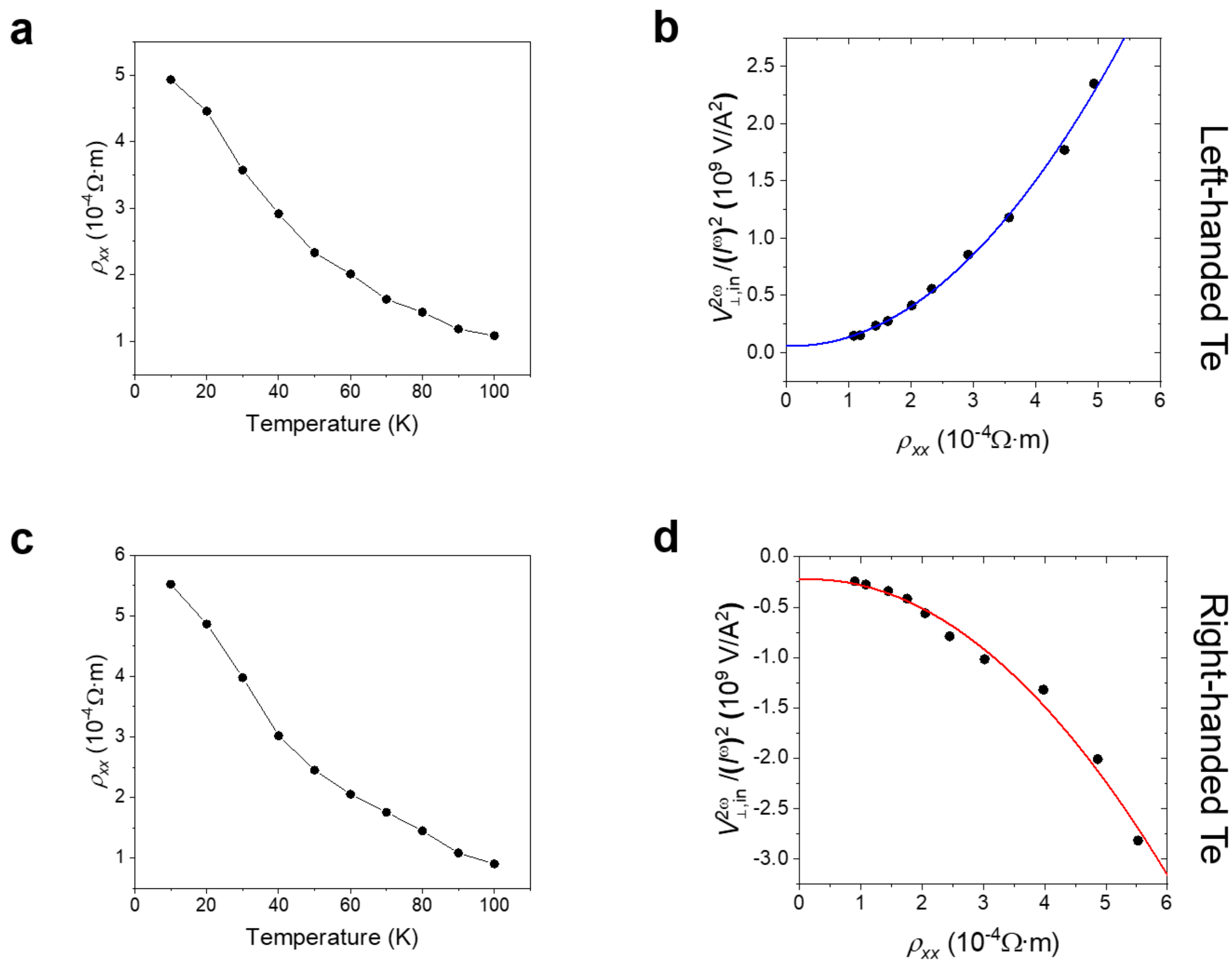

**Supplementary Figure 3. Scaling of cross-axial nonlinear Hall effect as a function of resistivity.** (a, c) Temperature dependent longitudinal resistivity $\rho_{xx}$ of a left- and right-handed Te device with the injected current $I^{\omega}$ at θ = 45°, respectively. (b, d) The ratio of the in-plane NHE voltages to the square of the input current (i.e. $V^{2\omega}_{\perp,in}/(I^{\omega})^2$) as a function of $\rho_l$ for the left- and right-handed Te device, respectively. The solid lines represent fits to the experimental data using the equation (5) in the main text.

| | $\chi_{xxx}$ |
|---|---|
| Left-handed Te Device (Device 1) | $2.31\times10^{-3}/\Omega\cdot V$ |
| Right-handed Te Device (Device 2) | $-2.22\times10^{-3}/\Omega\cdot V$ |

**Supplementary Table 1.** The nonlinear susceptibility $\chi_{xxx}$ of left- and right-handed Te devices.

| | $\beta$ | $\gamma$ | $\delta$ |
|---|---|---|---|
| In-plane, LH | 5.80 ($10^7$ V·A$^{-2}$) | -1.88 ($10^7$ m$^{-1}$·A$^{-1}$) | 9.52 ($10^{15}$ m$^{-2}$·V$^{-1}$) |
| In-plane, RH | -2.57 ($10^7$ V·A$^{-2}$) | 3.11 ($10^7$ m$^{-1}$·A$^{-1}$) | -8.68 ($10^{15}$ m$^{-2}$·V$^{-1}$) |

**Supplementary Table 2.** Fitted coefficients in equation (5) in the main text for scaling in-plane nonlinear Hall effect with longitudinal resistivity for each chirality.

**Supplementary Note 1: Symmetry analysis of the BCD**

Trigonal Te belongs to the space group $P3_121$ or $P3_221$. The BCD, defined as a rank-2 tensor $D_{ab}$, is constrained by the symmetries. The most general form of the BCD is written as

$$D = \begin{pmatrix} D_{xx} & D_{xy} & D_{xz} \\ D_{yx} & D_{yy} & D_{yz} \\ D_{zx} & D_{zy} & D_{zz} \end{pmatrix}$$

Under a symmetry operation $S$, The BCD tensor transforms as

$$D \to SDS^{-1}$$

Under the 3-fold screw rotation symmetry around the z axis $C_{3z}$, $D = C_{3z}DC_{3z}^{-1}$, $D_{xx} = D_{yy}$, $D_{xz} = D_{zx} = D_{yz} = D_{zy} = 0$ and $D_{xy} = -D_{yx}$. The constrained BCD form is given by

$$\begin{pmatrix} D_{xx} & D_{xy} & D_{xz} \\ D_{yx} & D_{yy} & D_{yz} \\ D_{zx} & D_{zy} & D_{zz} \end{pmatrix} \xrightarrow{C_{3z}} \begin{pmatrix} D_{xx} & D_{xy} & 0 \\ -D_{xy} & D_{xx} & 0 \\ 0 & 0 & D_{zz} \end{pmatrix}$$

Furthermore, the $D_{xy}$ component vanishes under the 2-fold rotation symmetry around the x-axis $C_{2x}$,

$$\begin{pmatrix} D_{xx} & D_{xy} & 0 \\ -D_{xy} & D_{xx} & 0 \\ 0 & 0 & D_{zz} \end{pmatrix} \xrightarrow{C_{2x}} \begin{pmatrix} D_{xx} & 0 & 0 \\ 0 & D_{xx} & 0 \\ 0 & 0 & D_{zz} \end{pmatrix}$$

Since the BCD is traceless, $D_{xx} = -D_{zz}/2$.

$$D_{xx} + D_{yy} + D_{zz} = \int \left(\partial_{k_x}\Omega_x + \partial_{k_y}\Omega_y + \partial_{k_z}\Omega_z\right) f_0 \; d^3k$$

$$= \int (\nabla \cdot \mathbf{\Omega}) f_0 \; d^3k = \int \nabla \cdot (\nabla \times \mathbf{A}) f_0 \; d^3k = 0$$

where **A** is a Berry connection. Consequently, the form of the BCD is as follows:

$$D = D_0 \begin{pmatrix} 1 & 0 & 0 \\ 0 & 1 & 0 \\ 0 & 0 & -2 \end{pmatrix}.$$

**Supplementary Note 2: Non-linear Susceptibility $\chi_{yxz}$ and $\chi_{xxx}$**

To extract each component of the non-linear susceptibility tensor of Te, we start from the non-linear electric field[1]

$$E^{2\omega} = \begin{pmatrix} \rho_{xx}\chi_{xxx}(E_x^{\omega})^2 \\ 2\rho_{yy}\chi_{yxz}E_x^{\omega}E_z^{\omega} \\ 0 \end{pmatrix} \tag{1}$$

For an in-plane current density $j^{\omega}$ applied at an angle $\theta$ in the x-z plane, the first-order electric field components are

$$E_x^{\omega} = \rho_{xx}j^{\omega}\sin\theta\,, E_z^{\omega} = \rho_{zz}j^{\omega}\cos\theta.$$

Using $V_{\perp,out}^{2\omega} = E_y^{2\omega}L_y$ and $j^{\omega} = I^{\omega}/(tw)$ for the out-of-plane transverse response, we obtain

$$\frac{V_{\perp,out}^{2\omega}}{(I^{\omega})^2} = 2\chi_{yxz}\rho_{xx}\rho_{yy}\rho_{zz}\frac{L_y}{t^2w^2}\sin\theta\cos\theta.$$

With $L_y = t,\ \rho_{ii} = R_{ii}A_{ii}/L_i,\ L_x = L_z$ and $A_x = A_z = tw$, this reduces to

$$\frac{V_{\perp,out}^{2\omega}}{(I^{\omega})^2} = 2\frac{R_{xx}R_{yy}R_{zz}}{{L_x}^2}A_y\chi_{yxz}\sin\theta\cos\theta.$$

Also, using $V_{\perp,in}^{2\omega} = E_x^{2\omega}cos\theta L_{\perp},\ \rho_{xx} = R_{xx}^{\omega}\cdot A_x/L_x$ , $L_{\perp} = L_x$ and $j^{\omega} = I^{\omega}/A_x$ for the in-plane transverse response, we have

$$\frac{V_{\perp,in}^{2\omega}}{(I^{\omega})^2} = \frac{1}{{L_x}^2}A_x(R_{xx}^{\omega})^3\chi_{xxx}\sin^2\theta\cos\theta.$$

**Supplementary Note 3: Calculation of the Berry Curvature Dipole.**

To estimate the Berry curvature dipole (BCD) contribution associated with the measured NHE, we start from the intrinsic BCD expression for the second-order non-linear susceptibility tensor with the relaxation time[2]

$$\chi_{abc} = -\varepsilon_{adc} \frac{e^3\tau}{2\hbar^2(1+i\omega\tau)} D_{bd}$$

where $\varepsilon_{adc}$ is the Levi-Civita symbol, $\tau$ is the scattering time, and $D_{bd}$ is the BCD tensor. Therefore, the intrinsic BCD contribution to the non-linear susceptibility tensor is given by:

$$\chi_{yxz} = \frac{e^3\tau}{2\hbar^2(1+i\omega\tau)} D_{xx}.$$

In the NHE, the out-of-plane Hall current is given by:

$$j_y^{2\omega} = \sigma_{yy} E_y^{2\omega} = 2\chi_{yxz} E_x^{\omega} E_z^{\omega}.$$

In the DC limit $\omega\tau \ll 1$, we have

$$\frac{E_y^{2\omega}}{E_x^{\omega} E_z^{\omega}} = \frac{1}{\sigma_{yy}} \frac{e^3\tau}{\hbar^2} D_{xx}.$$

**Supplementary Note 4: Scaling Law for the NHE of Te.**

To analyze the resistivity dependence of the $\chi_{abc}$ response in Te, we use the general multivariable scaling law of the non-linear Hall effect. The in-plane longitudinal resistivity along the current direction is

$$\rho_l(\theta) = \rho_{xx}\sin^2\theta + \rho_{zz}\cos^2\theta$$

Assuming one static and dynamic scattering channel, the scaling law can be written as[3]

$$\frac{V_y^{2\omega}}{(V_l^{\omega})^2} = \frac{1}{{\rho_l}^2}\left(C_1\rho_{l0} + C_2{\rho_{l0}}^2 + C_3\rho_{l0}\rho_{lT} + C_4{\rho_{lT}}^2\right)$$

with scaling parameters

$$C_1 = C^{sk,2}, C_2 = C^{in} + C_0^{sj} + C_{00}^{sk,1},$$

$$C_3 = 2C^{in} + C_0^{sj} + C_1^{sj} + C_{01}^{sk.1},$$

$$C_4 = C^{in} + C_1^{sj} + C_{11}^{sk,1}$$

where the disorder-independent coefficients for the intrinsic ($C^{in}$), side-jump ($C_1^{sj}$), intrinsic skew-scattering ($C_i^{sk.1}$), and extrinsic skew-scattering ($C_{ij}^{sk.2}$) contributions, respectively, and $\rho_{l0}$ is the residual longitudinal resistivity and $\rho_{lT} = \rho_l - \rho_{l0}$ is the temperature-dependent part.

The first-order in-plane longitudinal voltage is

$$V_l^{\omega} = \rho_l(\theta)\frac{L_x}{A_x}I^{\omega},$$

so that

$$\frac{V_y^{2\omega}}{(I^\omega)^2} = \left(\frac{L_x}{A_x}\right)^2 (C_1\rho_{l0} + C_2{\rho_{l0}}^2 + C_3\rho_{l0}\rho_{lT} + C_4{\rho_{lT}}^2).$$

Substituting $\rho_{lT} = \rho_l - \rho_{l0}$, we obtain

$$\frac{V_y^{2\omega}}{(I^\omega)^2} = \left(\frac{L_x}{A_x}\right)^2 [C_4{\rho_l}^2 + (C_3 - 2C_4)\rho_{l0}\rho_l + (C_1\rho_{l0} + (C_2 - C_3 + C_4){\rho_{l0}}^2)].$$

Absorbing the geometric factor into the fitting coefficients, the scaling law for Out-of-plane $V_{\perp,out}^{2\omega}$ becomes

$$\frac{V_{\perp,out}^{2\omega}}{(I^\omega)^2} = \beta + \gamma\rho_l + \delta{\rho_l}^2,$$

with

$$\beta = C_1\rho_{l0} + (C_2 - C_3 + C_4){\rho_{l0}}^2 = C^{sk,2}\rho_{l0} + \left(C_{00}^{sk.1} - C_{01}^{sk.1} + C_{11}^{sk.1}\right){\rho_{l0}}^2,$$

$$\gamma = (C_3 - 2C_4)\rho_{l0} = \left(C_0^{sj} - C_1^{sj} + C_{01}^{sk,1} - 2C_{11}^{sk,1}\right)\rho_{l0},$$

$$\delta = C_4 = C^{in} + C_1^{sj} + C_{11}^{sk,1}.$$

For the *In-plane* $V_{\perp,in}^{2\omega}$, only the x-component of the applied electric field contributes, yielding $\rho_l(\theta) = \rho_{xx}$ by eq.1. Following the same steps, the scaling law for the *In-plane* $V_{\perp,in}^{2\omega}$ becomes

$$\frac{V_{\perp,in}^{2\omega}}{(I^\omega)^2} = \beta + \gamma\rho_{xx} + \delta{\rho_{xx}}^2,$$

with

$$\beta = C_1\rho_{l0} + (C_2 - C_3 + C_4){\rho_{l0}}^2 = C^{sk,2}\rho_{l0} + \left(C_{00}^{sk.1} - C_{01}^{sk.1} + C_{11}^{sk.1}\right){\rho_{l0}}^2,$$

$$\gamma = (C_3 - 2C_4)\rho_{l0} = \left(C_0^{sj} - C_1^{sj} + C_{01}^{sk,1} - 2C_{11}^{sk,1}\right)\rho_{l0},$$

$$\delta = C_4 = C_1^{sj} + C_{11}^{sk,1}.$$

Note that the *In-plane* $V_{\perp,in}^{2\omega}$ cannot have an intrinsic origin (i.e. $C^{in} = 0$), since the $\chi_{xxx}$ is off-diagonal.